\documentclass{emulateapj}
\newcommand{\mbh}{M$_{\rm BH}$}
\newcommand{\msigma}{M$_{\rm BH}$-$\sigma_{*}$}

\shorttitle{The \msigma\ relation in the last six billion years}
\shortauthors{Woo et al.}

\begin{document}

\title{Cosmic Evolution of Black Holes and Spheroids. III. The \msigma\ relation in the last six billion years}

\author{Jong-Hak Woo\altaffilmark{1}, Tommaso Treu\altaffilmark{1,2},
Matthew A. Malkan\altaffilmark{3}, Roger D. Blandford\altaffilmark{4}}

\altaffiltext{1}{Department of Physics, University of California,
Santa Barbara, CA 93106-9530; woo@physics.ucsb.edu,
tt@physics.ucsb.edu} 
\altaffiltext{2}{Sloan Fellow, Packard Fellow}
\altaffiltext{3}{Department of Physics and Astronomy, University of California at Los Angeles, CA 90095-1547, malkan@astro.ucla.edu} \altaffiltext{4}{Kavli Institute for Particle Astrophysics and Cosmology, Stanford, CA, rdb@slac.stanford.edu}

\begin{abstract}

We measure the evolution of the correlation between black hole mass
and host spheroid velocity dispersion (\msigma) over the last 6
billion years, by studying three carefully selected samples of active
galaxies at $z=0.57$, $z=0.36$ and $z<0.1$. For all three samples,
virial black hole masses are consistently estimated using the line
dispersion of H$\beta$ and the continuum luminosity at 5100\AA~or
H$\alpha$ line luminosity, based on our cross calibration of the broad
line region size-luminosity relation.  For the $z=0.57$ sample, new
stellar velocity dispersions are measured from high signal-to-noise
ratio spectra obtained at the Keck Telescope, while for the two lower
redshift samples they are compiled from previous works. Extending our
previous result at $z=0.36$, we find an offset from the local
relation, suggesting that for fixed M$_{\rm BH}$, distant spheroids
have on average smaller velocity dispersions than local ones.  The
measured offset at $z=0.57$ is $\Delta \log \sigma_{*}=0.12 \pm
0.05\pm 0.06$ (or $\Delta \log M_{\rm BH}=0.50 \pm 0.22\pm 0.25$),
i.e. $\Delta \log M_{\rm BH} = (3.1\pm1.5)\log (1+z) +
0.05\pm0.21$. This is inconsistent with a tight and non-evolving
universal \msigma\, relation at the 95\%CL.

\end{abstract}


\section{Introduction}

Understanding the origin of the black hole mass - spheroid velocity
dispersion (\msigma) relation (Ferrarese \& Merritt 2000; Gebhardt et
al.\ 2000) is a key goal of unified models of black hole -- galaxy
coevolution (e.g. Kauffmann \& Haenhelt 2000; di Matteo et al.\ 2005;
Ciotti \& Ostriker 2007).  One of the most powerful observational
tests of the proposed explanations is to measure the time evolution of
the \msigma\ relation since various scenarios predict different cosmic
evolution.  For example, -- for a fixed \mbh\ -- Robertson et al.\
(2006) predict an increase of $\sigma_{*}$ with redshift, Croton
(2006) and Bower et al. (2006) predict a decrease, while Granato et
al. (2004) expect no evolution.

In recent years, a number of groups have investigated the evolution of
the \msigma\ relation, using various techniques to estimate $\sigma_*$
of AGN host galaxies (e.g. Shields et al. 2003; Walter et al.\ 2004;
Salviander et al.\ 2007). Starting with our pilot study (Treu et
al. 2004), we reported the first direct measurement of the \msigma\
relation beyond the local Universe (Woo et al. 2006, hereafter paper
I), and updated it with corrected AGN continuum luminosities using
Hubble Space Telescope (HST) images in paper II (Treu et al. 2007).

By observing 14 Seyfert 1 galaxies, we determined stellar velocity
dispersions in the integrated spectra, and \mbh\ from AGN broad
emission line widths, which are thought to measure the gravity of the
central mass on sub-parsec scales. We found that the measured \msigma\
relation at $z=0.36$ is offset with respect to the local relationship
($\Delta \log$ \mbh = 0.54$\pm$0.12$\pm$0.21 at fixed $\sigma_{*}$).
In other words black holes of a fixed mass appeared to live
in bulges with smaller velocity dispersion 4 Gyrs ago (at 95\% CL), consistent
with recent growth and evolution of intermediate mass spheroids. Using
HST images, we obtained a consistent result, $\Delta \log {\rm M}_{\rm
BH}> 0.51\pm0.14\pm0.17$, by measuring the M$_{\rm BH}$- spheroid
luminosity relation of the same sample (paper II).  This result may be
consistent with a scenario where intermediate-mass blue galaxies
undergo merging at relatively recent times and arrive on the local
black hole-galaxy relations by becoming more massive red
galaxies. However, much work remains to be done due to the small sample size
and large uncertainties, before this initial result can
become a high precision measurement.

We report here our first measurement at the next redshift window
($z=0.57$, adding $\sim$50\% to the look-back time), so that
evolutionary trends can be measured over a longer range in cosmic
time. We also improve the local baseline by consistently estimating
\mbh\, for a sample of 48 nearby Seyfert 1 galaxies with published
stellar velocity dispersion (Greene \& Ho 2006). To minimize
repetition, readers are referred to our previous works (papers I, II;
McGill et al.\ 2008; hereafter M08) for detailed discussions of the
systematics inherent to the measurement. The paper is organized as
follows. Section 2 describes sample selection and
observations. Section 3 presents our measurements.  Section 4 presents
the \msigma\ relation.  Discussion and conclusions are presented in
\S~5.  We adopt $\Omega_m=0.3$, $\Omega_{\Lambda}=0.7$, and $H_{0}=70$
km sec$^{-1}$ Mpc$^{-1}$.

\section{Data}

\begin{figure}
\epsscale{0.8}
\plotone{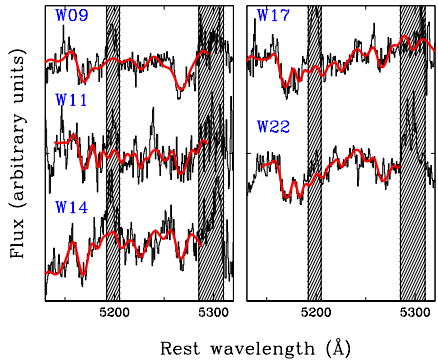}
\caption{Velocity dispersion measurements. The region including the
main stellar features is shown together with the best fit template
(red thick line). The regions around narrow AGN emission lines --
identified by vertical lines -- are masked out before fitting.}
\label{fig_dis}       
\end{figure}

A sample of broad-line AGNs was selected from the Sloan Digital Sky
Survey Data Release 4 (SDSS DR4).  Following our strategy at
$z=0.36\pm0.01$ (paper I), we chose the next redshift window,
$z=0.57\pm0.01$, to avoid strong sky features on the redshifted
stellar lines around the Mg-Fe line region, minimizing the
uncertainties related to sky subtraction and atmospheric absorption
corrections.

Our selection procedure was slightly modified with respect to that of
the lower redshift sample. Initially, 365 broad-line AGNs at
z=0.57$\pm$0.01 were collected from SDSS DR4, based on the presence of
the broad H$\beta$ line. Out of 365 AGNs, we selected 20 objects with
g'-r'$>$0.1 and r'-i'$>$0.3 (AB), expecting non-negligible stellar
light in the observed spectra, based on stellar and AGN spectral
models. The effects of this color cut will be modeled in detail in
future papers, when Keck and possibly HST data for a larger sample at
$z=0.57$ will be available. However, since the colors of the new
sample are similar to those of the $z=0.36$ sample, we do not
expect the color cut to introduce a significant bias. In any case, the color
cut will tend to select more massive host galaxies for a given
nuclear luminosity. Hence, if any biased is introduced, it
should bias against the offset seen in papers I and II.

High signal-to-noise (S/N) ratio spectra for 5 objects were obtained with
the LRIS spectrograph (Oke et al. 1995) at the Keck-I telescope during
two runs in January 2007 and April 2007. The 831 lines mm$^{-1}$
grating centered at 7600\AA~was used with a $1"$ wide slit, yielding a
pixel scale of 0.92\AA$\times 0\farcs$215 and a Gaussian resolution ($\sigma$)
$\sim$58 km s$^{-1}$. Observing conditions were generally favorable
with 0.7-1.2" seeing. Total exposure times for each object ranges
between 2.5 and 3.5 hours.  The observing strategy, data reduction,
calibration, and one-dimensional spectra extraction processes were
very similar to those described in paper I.

\section{Measurements}

\begin{deluxetable*}{lcccccccccc}
\tablewidth{0pt}
\tablecaption{Targets and Measured Properties}
\tablehead{
\colhead{Name}        &
\colhead{z}           &
\colhead{RA (J2000)}     &
\colhead{DEC (J2000)}    &
\colhead{$i'$}    &
\colhead{Exp.}       &
\colhead{S/N}        &  
\colhead{$\sigma_{H\beta}$} &
\colhead{$\lambda L_{5100}$}    &
\colhead{log M$_{\rm BH}/M_{\odot}$}  &
\colhead{$\sigma_{*}$}    \\
\colhead{} &
\colhead{} &
\colhead{} &
\colhead{} &
\colhead{mag} &
\colhead{hr} &
\colhead{\AA$^{-1}$} &
\colhead{km s$^{-1}$ } &
\colhead{$10^{44}$ erg s$^{-1}$} &
\colhead{}      &
\colhead{km s$^{-1}$}      \\
\colhead{(1)} &
\colhead{(2)} &
\colhead{(3)} &
\colhead{(4)} &
\colhead{(5)} &
\colhead{(6)} &
\colhead{(7)} &
\colhead{(8)} &
\colhead{(9)} &
\colhead{(10)} &
\colhead{(11)} }
\tablecolumns{11}
\startdata
W9 & 0.5651       &  15 52 27.82 &+56 22 36.47 & 19.00 & 2.5  & 79  &   2598  &  4.31 &   8.64  & 289$\pm $   19  \\
W11& 0.5649       &  1 55 16.18  &-9 45 55.99  & 20.03 & 3    & 32  &   2103  &  1.53 &   8.15  & 126$\pm $   21  \\
W14& 0.5616       &  12 56 31.90 &-2 31 30.62  & 18.71 & 2.5  & 94  &   2192  &  4.94 &   8.54  & 228$\pm $   20  \\
W17& 0.5611       &  10 07 28.38 &+39 26 51.83 & 19.71 & 2.5  & 32  &   2320  &  2.00 &   8.31  & 165$\pm $   14  \\
W22& 0.5649       &  3 42 29.70  &-5 23 19.49  & 18.60 & 3.5  & 101 &   2442  &  5.77 &   8.68  & 144$\pm $   21  \\
\enddata
\label{T_data}
\tablecomments{
Col. (1): Target ID.  
Col. (2): Redshift from SDSS-DR6.
Col. (3): Right Ascension.
Col. (4): Declination.
Col. (5): Extinction corrected $i'$ AB magnitude from SDSS photometry.
Col. (6): Total exposure time.
Col. (7): Signal-to-noise ratio of the combined spectrum (average in the 8000-8300\AA spectral region).
Col. (8): Second moment of H$\beta$ in km s$^{-1}$. Typical error is $\sim$10\%.
Col. (9): Rest frame luminosity at 5100\AA.  Typical error is a few \%.
Col. (10): Logarithm of \mbh~ in solar units. Estimated uncertainty is 0.4 dex.
Col. (11): Stellar velocity dispersion.}
\end{deluxetable*}

This section describes our measurement of $\sigma_*$ and \mbh\ for the
5 Seyferts at $z=0.57$ (\S~\ref{ssec:sigma} and \S~\ref{ssec:bhm}),
and \mbh\ estimates for the 48 local Seyferts (\S~\ref{ssec:local}).
The relevant properties of the $z=0.57$ sample are listed in
Table~\ref{T_data}.

\subsection{Stellar Velocity Dispersion}
\label{ssec:sigma}

We used the Mg-Fe region (rest-frame $\sim$ 5050-5300\AA) to measure
velocity dispersion as described in detail in paper I.  Here, we
briefly summarize the procedure and systematic uncertainties.
First, we subtracted broad AGN Fe emission, using a set of I Zw 1
templates.  Then, we compared in pixel space the observed spectra with
5 stellar templates (G8, G9, K0, K2, and K5 giant) broadened with a
range of Gaussian velocity.  AGN narrow emission lines (e.g. [N I]
5201\AA~ and [Fe XIV] 5304\AA) were masked out before fitting, as
shown in Figure~1.  Fits were performed for all templates to estimate
the effect of template mismatch, yielding comparable measurements
within the errors (10-20\%).  The best-fit template was used for the
final dispersion measurements.

The Mg-Fe region typically used for dynamical studies is a natural
choice for our sample since other strong stellar features such as the
CaII triplet are out of the optical spectral range.  Feature mismatch
due to $\alpha$-enhancement in massive early-type galaxies is a
well-known problem in kinematics studies (e.g. Barth et al. 2003; Woo
et al. 2004) and can potentially increase systematic uncertainties.
However, in paper I we found that only one out of 14 Seyfert galaxies
at $z=0.36$ shows signs of Mg mismatch, as expected because the
inferred stellar velocity dispersions are more typical of a Milky Way
type galaxy than of a massive early-type galaxy. As for the lower
redshift sample, we do not find significant mismatch in our $z=0.57$
sample, as shown in Figure 1.

Following the procedure described in paper I, we estimate a total
systematic uncertainty of 0.05 dex on $\sigma$, combining the effects
of template mismatch, potential errors due to the large spectroscopic
aperture, and host galaxy morphology and inclination.  This translates
into 0.20 dex uncertainty of the offset in $\log$ \mbh\ from the
\msigma\ relation.

\subsection{Black Hole Mass}
\label{ssec:bhm}

Black hole mass can be estimated using the `virial' method based on the
empirical relation between the size of the broad line region and
continuum luminosity of the reverberation sample (Kaspi et al.\
2005), and the velocity scale given by the width of the broad emission lines.
In practice, we measured the line dispersion of broad H$\beta$ by
fitting the observed line profile with Gauss-Hermite polynomials as
described in paper I and in M08. The continuum luminosity around
5100\AA\ (L$_{\rm 5100}$) was measured by averaging flux in the 5070-5130\AA\,
region. Considering the difficulty of achieving absolute flux
calibration for the Keck spectra -- due to slit losses, variable
seeing and sky transparency-- we re-calibrated our spectrophotometry
with the extinction corrected $i'$ band magnitude taken from the
SDSS-DR6 archive, by calculating and correcting for the offset between
Sloan and our synthetic $i'$ band magnitude measured from the observed
spectra.

For low luminosity AGNs (L$_{5100} < 10^{44}$ erg s$^{-1}$) continuum
luminosity can be overestimated due to the significant contribution
from host galaxies. Thus, correcting for the host galaxy contamination
is crucial to avoid overestimation of \mbh.  The size-luminosity
relation was in fact revised with a lower slope ($\sim$0.5 as expected
in photoionization scenarios) and a higher normalization, after
correcting for the galaxy contamination in low luminosity AGNs in the
reverberation sample (Bentz et al. 2006a).

It requires high resolution HST imaging
to correct for the host galaxy contamination for distant AGNs.
Since this is not available for
our sample at the moment, we cannot but adopt the size-luminosity
relation based on the total (observed) luminosity.  However, based on
our experience at $z=0.36$, host galaxy contamination is not expected
to be a major effect. In paper II, for Seyfert galaxies with similar
luminosity, we compared \mbh\ estimates based on the size-luminosity
relation of Kaspi et al.\ (2005) with new estimates based on the
revised size-luminosity relations of Bentz et al. (2006a), after
correcting for host galaxy contamination using HST images.  We found
that new \mbh\ estimates are on average 0.09 dex smaller, due to the
combined effects of removing host galaxy light while using the new
size-luminosity relation with a higher normalization.

Therefore, we will adopt as our best estimate of \mbh, the following
equation from Paper I based on Kaspi et al. (2005) and Onken et
al. (2004), equivalent to the most recent calibration of empirical
\mbh\ estimators from M08:
\begin{equation}
M_{\rm BH} = 10^{8.33}{\rm M}_{\odot} \times \left(
{\sigma_{H\beta} \over 3000 {\rm km s}^{-1} } \right)^{2} \left( {\lambda
L_{5100} \over 10^{44} {\rm erg s}^{-1}} \right)^{0.69}~,
\label{eq_BHM} 
\end{equation}
where $\sigma_{\rm H\beta}$ is the line dispersion (second moment) of
H$\beta$. We assume 0.4 dex uncertainty on the estimated \mbh, based
on comparisons of reverberation data and single-epoch data
(Vestergaard \& Peterson 2006; M08), which dominates the errors on
$\sigma_{\rm H\beta}$ and L$_{\rm 5100}$.

As a sanity check, we compared \mbh\ estimates based on Equation 1
with those based on the new size-luminosity relation (Bentz et
al. 2006a) along with the same virial coefficient of Onken et
al. (2004):
\begin{equation}
M_{\rm BH} = 10^{8.58}{\rm M}_{\odot} \times \left(
{\sigma_{H\beta} \over 3000 {\rm km s}^{-1} } \right)^{2} \left( {\lambda
L_{5100,n} \over 10^{44} {\rm erg s}^{-1}} \right)^{0.518}~,
\label{eq_BHM2}
\end{equation}
where L$_{5100,n}$ is the {\it nuclear} luminosity at 5100\AA\ after
correcting for host galaxy contamination.  Since high resolution
images needed for an accurate measurement of the nuclear luminosity
are not available for our sample, we assume an average AGN fraction in
the observed light at 5100\AA.  
If the host galaxy contamination is negligible (L$_{\rm 5100, n}$=
L$_{\rm 5100}$), Equation 2 gives 0.16 dex higher \mbh\ compared to Equation 1,
while if the AGN fraction is assumed to be 50\%, \mbh\ is 0.006 dex higher. 
Thus, using Equation 1 without correcting for the host galaxy contamination --
which we cannot do at the moment -- does not significantly affect 
our \mbh estimates.   
As in paper I, we adopt a systematic error of 0.11 dex in \mbh\ 
estimates, which is dominated by AGN continuum luminosity uncertainty
due to host galaxy contamination.

\subsection{Local Seyferts}
\label{ssec:local}

To measure the evolution of the \msigma\ relation, it is important to
have a well defined local sample. The sample of 14 Seyfert galaxies
with reverberation \mbh, and measured stellar velocity dispersion
(Onken et al. 2004) is a good local benchmark. However, it is
desirable to have a complementary sample for two reasons.  First, the
reverberation sample is small in size and shows a flattened
distribution on the \msigma\ plane, especially with a new
reverberation black hole mass of NGC 4151 (Bentz et al. 2006b; see
magenta points in Figure 2). Second, there could be an unknown
systematic offset between the reverberation mass and our single-epoch
mass due to the uncertainties in measuring velocity and luminosity
from single-epoch spectra, potentially caused by, e.g., flux
variability, velocity variability, the narrow line subtraction
(e.g. Collin et al. 2006; Woo 2008).

For these reasons, we estimated \mbh\ for a sample of local Seyferts,
using the line dispersion of H$\beta$ and H$\alpha$ line luminosity
(L$_{{\rm H}\alpha}$), and a formula consistently calibrated 
with that used for \mbh\ estimates at $z=0.36$ and $z=0.57$ (M08). 
We selected 55 Seyfert 1 galaxies at $z < 0.1$ from SDSS-DR6,
with published stellar velocity dispersion (Greene \& Ho 2006). Seven objects were
excluded due to the very faint broad component of H$\beta$ that
prevented us from measuring reliable line widths. For local low
luminosity Seyferts, host galaxy light is a significant fraction of the
light observed within the Sloan fiber ($3"$ diameter), superimposing
strong stellar absorption on the broad H$\beta$ line profile. Thus, we
subtracted the stellar features, using eigenspectra templates 
based on a principal component analysis of several hundred galaxy spectra 
(Hao et al. 2005). Since the L$_{5100}$ measured from SDSS spectra could be also
significantly contaminated by stellar light, we used L$_{{\rm
H}\alpha}$ from Greene \& Ho (2006) instead, together with the \mbh\ recipe
calibrated by M08 and Green \& Ho (2005).

\begin{figure*}
\epsscale{1.1}
\plottwo{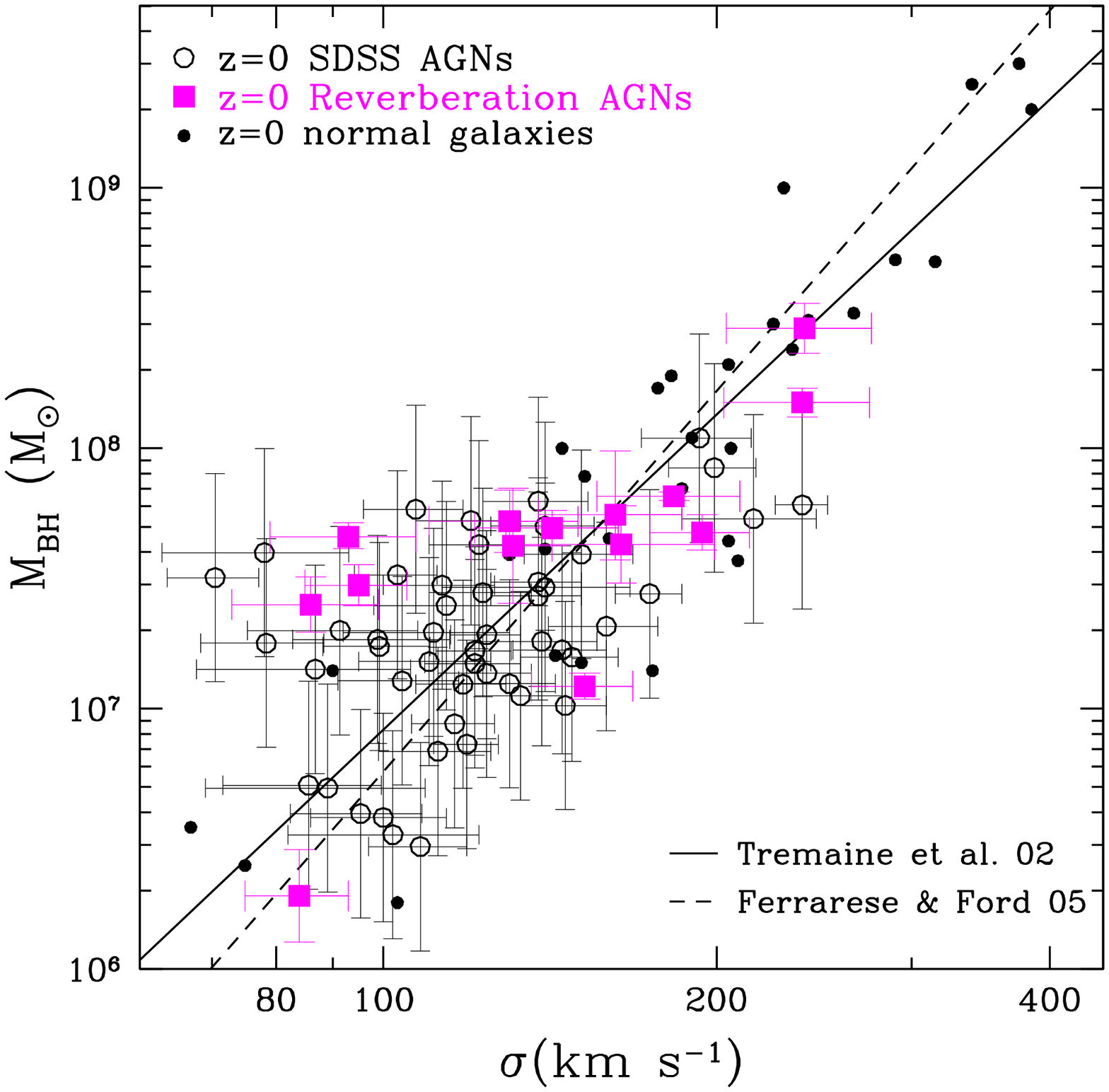}{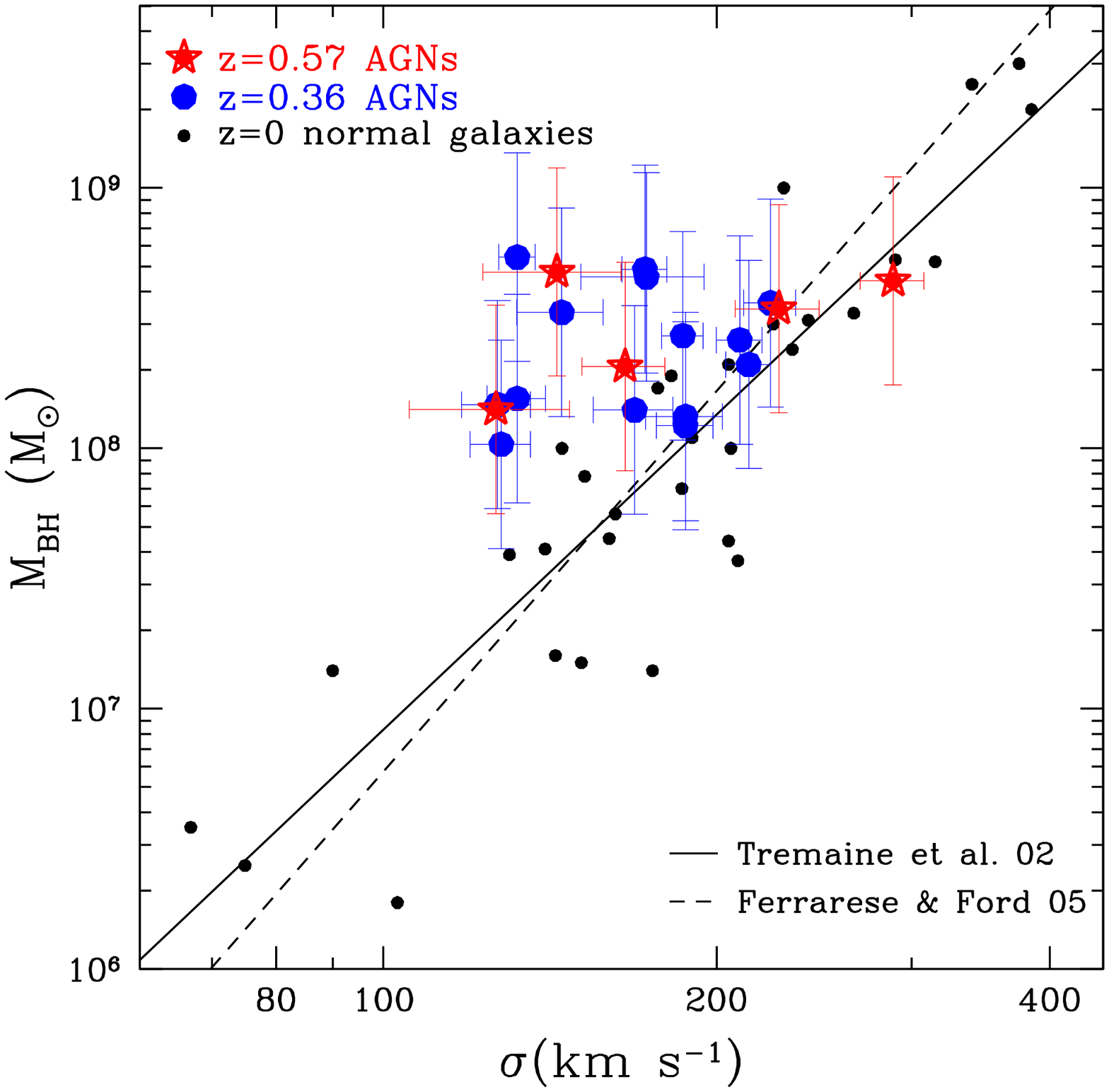}
\caption{The \msigma\ relation of active galaxies. Left panel: 
local Seyferts with $\sigma_{*}$ from Greene \& Ho (2006) and our own M$_{\rm BH}$ estimates, 
consistently calibrated with our estimates for distant samples (black circles);
local Seyferts with \mbh, measured via reverberation mapping (Onken et
al. 2004; magenta squares). Right panel: new measurements at $z=0.57$ (red stars);
Seyfert galaxies at $z=0.36$ from our earlier work (paper I, II; blue circles). 
The local relationships of quiescent
galaxies (Tremaine et al. 2002; black points) are shown for comparison
as a solid (Tremaine et al. 2002) and dashed (Ferrarese \& Ford 2005)
line.}
\label{fig_all}
\end{figure*}

\section{The \msigma\ relation}
\label{sec:msigma}

\begin{figure}
\epsscale{1}
\plotone{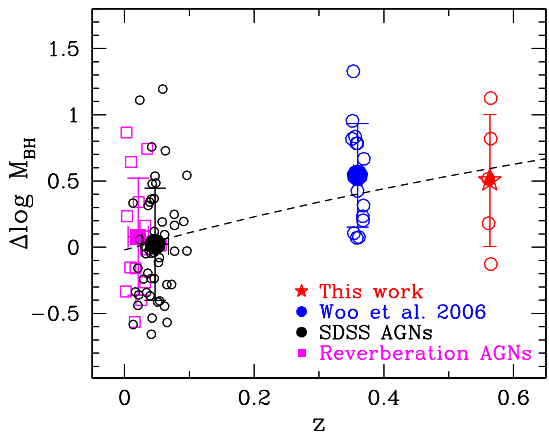}
\caption{Offset in M$_{\rm BH}$ with respect to the local quiescent
sample (Tremaine et al.\ 2002) as a function of redshift.  Large solid
points with error bars represent the average and rms scatter for the
four samples of active galaxies. The rms scatter of the $z=0.57$
sample is 0.5 dex, similar to that of local active galaxies.  Note
that all 'virial' \mbh\ are based on the second moment of H$\beta$ and
the same calibration of the virial coefficient. The dashed line
represent the best fit relation $\Delta \log M_{\rm BH} =
(3.1\pm1.5)\log (1+z) + 0.05\pm0.21$.}
\label{fig_evolution}
\end{figure}
In Figure~\ref{fig_all}, the \msigma\ relation for local active 
galaxies (left panel) and our samples at $z=0.36$ and $z=0.57$ (right panel)
are presented along with local quiescent galaxies.
Two local AGN samples (SDSS sample from \S~3.3 and the reverberation
sample from Onken et al. 2004) are consistent with the \msigma\
relation of quiescent galaxies, although the scatter is somewhat
larger (r.m.s. 0.45 and 0.43 dex, respectively for the SDSS sample and
the reverberation sample) compared to that of quiescent galaxies
($\sim$0.3 dex).  The scatter increases as galaxy mass decreases,
perhaps consistent with mass-dependent evolution in the sense that
less massive galaxies are still evolving to the \msigma\
relation. This may indicate that the \msigma\ relation is not as tight
for late-type galaxies even at $z\sim0$. Splitting evenly the local
sample into two groups below and above $\sigma_{*}=120$km s$^{-1}$,
and taking into account the measurement errors on $\sigma_{*}$, we
find that the intrinsic scatter is a factor of 2 larger for the low
$\sigma_*$ sample (0.43 vs 0.22).

The distant samples are offset from the local \msigma\ relation.  The
average offset of the $z=0.57$ sample is $0.50\pm0.22\pm0.25$ dex in
\mbh, corresponding to $0.12\pm0.05\pm0.06$ in $\log \sigma_{*}$ -- in
the sense that velocity dispersions were on average smaller for given
\mbh\ six Gyrs ago (Figure~\ref{fig_evolution}). Using the new
size-luminosity relation of Bentz et al. 2006a (Equation 2) and
assuming an average AGN fraction $\sim$50\%, we find an equivalent
offset, $\Delta$log\mbh = 0.51. If the AGN fraction is higher, then
the offset increases (see Section 3.2), indicating that \mbh\
estimates based on Equation 1 is not significantly overestimated. The
result is similar to the average offset of the $z=0.36$ sample (papers
I and II), although the error bars on the measurement are large enough
to allow for a variety of redshift trends.  We include in our error
analysis, in addition to the random errors, a potential systematic
error of 0.25 dex, estimated by combining systematic uncertainties in
\mbh\ and $\sigma_*$.

To quantify the significance of evolution, we consider the three
active samples. We emphasize that \mbh\, was consistently estimated
based on the line dispersion of H$\beta$ and the same virial
coefficient (shape factor).  Thus, a change in the virial coefficient
will move all samples vertically by the same amount, keeping the
offset unchanged, unless the kinematics of the broad line region
(hence, the virial coefficient) varies as a function of \mbh\ or
redshift.  Therefore, we consider the systematic error on the relative
calibration of \mbh\ to be negligible, leaving systematic errors in
the measurement of $\sigma_*$ as the main source of systematic
uncertainty in the evolution.  Including random and systematic errors
in the analysis, we find that the best fit relation is $\Delta \log
M_{\rm BH} = (3.1\pm1.5) log (1+z) + (0.05\pm0.21)$, i.e. the slope is
non zero at the two sigma level.  However, as discussed in paper I and
II, it is important to keep in mind that the observed offset may not
represent evolution, if the higher z samples are not direct
progenitors of the lower z samples due to, e.g., the somewhat
different scales in galaxy mass and \mbh.

\section{Discussion and conclusions}

We investigated the evolution of the \msigma\ relation using three
samples of Seyfert galaxies at $z<0.1$, $z=0.36$, and $z=0.57$,
finding evolution in the last 6 Gyr at the 95\%CL.  This result is
consistent with a scenario where black hole growth predates bulge
assembly and that bulges grow substantially in the last 6 Gyr -- at
least at this mass scale -- if the local \msigma\ relation is the
universal end-point of black hole-galaxy coevolution.

As discussed in paper I, collisional merging of late-type galaxies can
drive the evolution of the \msigma\ relation. The mass and stellar
velocity dispersion of the final spheroid will increase, not only by
forming new stars but also transforming rotation-supported disk stars
into pressure-supported spheroid components. This can potentially
overcome the growth in \mbh\ due to merging with the supermassive
black hole of the companion, especially if the companion galaxy is 
not spheroid dominated. 


In a galaxy merging scenario, the evolution of the \msigma\ relation
could be mass-dependent, similarly to the downsizing trends in galaxy
evolution (Cowie et al.\ 1996) and AGN evolution (Barger et al.\
2005).  As seen for example in fundamental plane studies (e.g. Treu et
al. 2005, Woo et al.\ 2004, 2005), active and quiescent massive
early-type galaxies have relatively old stellar populations in the
redshift range considered here ($z\sim0.4-0.6$). Together with the
results on the evolution of the mass function (e.g. Bundy et
al. 2007), this is consistent with an early epoch of assembly for the
most massive spheroids. Thus, the evolution of the \msigma\ relation
could be mass dependent, slower at this redshift for the more massive
galaxies (see Peng et al. 2006 for the \mbh--spheroid luminosity
relation of massive high redshift galaxies, which show evolution in
the same sense as our sample since z$\sim$2).  Recently, Shen et
al. (2008) present the \msigma relation out to $z\sim0.4$
based on SDSS spectra, concluding that the offset (in the same direction as
the one reported here) with redshift is not significant for their
sample. However, since their 28 galaxies with measured $\sigma_*$ at
z$>$0.3 have an average S/N$=$18.7 per pixel (and hence the S/N of the
stellar spectrum is less than $\sim$10 per pixel if the nuclear light
fraction is $\sim$50\%) and \mbh\ was based on the FWHM of H$\beta$
line, direct comparison with our result is not straightforward.  As
discussed in paper II, the broad observational picture is far from
conclusive at the moment, requiring larger samples over a wider mass
range than the present sample to test this hypothesis.

It is important to consider selection effects.  First, since our
samples were selected based on the flux and width of the broad lines,
they could be biased towards high \mbh~ objects (paper II; see also
Lauer et al. 2007b). However, as we showed in paper II with Monte
Carlo simulations, this bias is too small to account for the observed
offset\footnote{The \mbh\ range and measured offset are similar to
those of the sample studied in paper II, resulting in the same
negligible bias $<0.1$ dex.}, unless the intrinsic scatter of the
\mbh-$\sigma_*$ relation at $z=0.57$ -- which is unknown -- is of
order 1 dex. Second, although active galaxies are the only target for
\mbh\ estimation in the distant universe, they may not represent the
general galaxy population, as they are rare objects with a highly
accreting and radiatively efficient black hole. However, two pieces of
evidence argue against the explanation of the observed evolution
purely in terms of systematic differences between active and quiescent
galaxies: i) a consistent \msigma\ relation is found locally for the
two active galaxy samples; ii) the \msigma\ of distant active galaxies
is offset from that of the local active sample.

An alternative or complementary explanation of the observed offset is
that the \msigma\ relation is not tight for late-type galaxies, as
perhaps suggested by the increasing scatter for local active samples,
especially at the low mass end. This scenario is consistent with the
idea of downsizing, with low mass blue late type-galaxies yet to join
the more massive red early-type galaxies on the tight \msigma\
relation.  So far, only a few late-type galaxies are included in the
local quiescent galaxy sample that defines the local
\msigma\ relation.  A larger sample with more disk-dominant quiescent
galaxies is needed to investigate any systematic difference in the
local scaling relations.


\acknowledgments

This work is based on data collected at the Keck Observatory, operated
by Caltech, UC, and NASA, and is made possible by the public archive
of the Sloan Digital Sky Survey. T.T.  acknowledges support from the
NSF through CAREER award NSF-0642621, from the Sloan Foundation, and
from the Packard Foundation. We acknowledge financial support from NASA
through HST grant AR-10986. We thank C.~Peng and M.~Favata for useful
discussions, L.~Hao for providing the PCA algorithm, and the referee 
for useful suggestions.


\clearpage


\begin{thebibliography}{}
\bibitem[Barger et al.(2005)]{2005AJ....129..578B} Barger, A.~J., Cowie, 
L.~L., Mushotzky, R.~F., Yang, Y., Wang, W.-H., Steffen, A.~T., \& Capak, 
P.\ 2005, \aj, 129, 578 

\bibitem[Bentz et al.(2006)]{2006ApJ...644..133B} Bentz, M.~C., Peterson, B.~M., Pogge, R.~W., Vestergaard, M., \& Onken, C.~A.\ 2006a, \apj, 644, 133 

\bibitem[Bentz et al.(2006)]{2006ApJ...651..775B} Bentz, M.~C., et al.\ 2006b, \apj, 651, 775

\bibitem[Bower et al.(2006)]{2006MNRAS.370..645B} Bower, R.~G., et al. \ 2006, \mnras, 370, 645 

\bibitem[Bundy et al.(2007)]{2007ApJ...665L...5B} Bundy, K., Treu, T., \& 
Ellis, R.~S.\ 2007, \apjl, 665, L5

\bibitem[Ciotti \& Ostriker(2007)]{2007ApJ...665.1038C} Ciotti, L., \& 
Ostriker, J.~P.\ 2007, \apj, 665, 1038

\bibitem[Collin et al.(2006)]{2006A&A...456...75C} Collin, S., Kawaguchi, T., Peterson, B.~M., \& Vestergaard, M.\ 2006, \aap, 456, 75 

\bibitem[Cowie 1996]{Cowie96} Cowie, L. L., Songaila, A., Hu, E. M., \& Cohen J. G. 1996, AJ, 112, 839
\bibitem[Croton(2006)]{2006MNRAS.369.1808C} Croton, D.~J.\ 2006, \mnras, 369, 1808
\bibitem[Di Matteo et al.(2005)]{2005Natur.433..604D} Di Matteo, T., 
Springel, V., \& Hernquist, L.\ 2005, \nat, 433, 604 

\bibitem[Ferrarese \& Merritt(2000)]{2000ApJ...539L...9F} Ferrarese, L., \& Merritt, D.\ 2000, \apjl, 539, L9 

\bibitem[{{Ferrarese} \& {Ford}(2005)}]{F+F05}
{Ferrarese}, L. \& {Ford}, H. 2005, Space Science Reviews, 116, 523

\bibitem[Gebhardt et al.(2000)]{2000ApJ...539L..13G} Gebhardt, K., et al.\ 2000, \apjl, 539, L13
\bibitem[Granato et al.(2004)]{2004ApJ...600..580G} Granato, G.~L., De Zotti, G., Silva, L., Bressan, A., \& Danese, L.\ 2004, \apj, 600, 580
\bibitem[Greene \& Ho(2005)]{2005ApJ...630L..122G} Greene, J.~E., \& Ho, L.~C.\ 2005, \apj, 630, 122
\bibitem[Greene \& Ho(2006)]{2006ApJ...641L..21G} Greene, J.~E., \& Ho, L.~C.\ 2006, \apjl, 641, L21
\bibitem[Hao et al.(2005)]{2005AJ....129.1795H} Hao, L., et al.\ 2005, \aj, 129, 1795 

\bibitem[Hopkins et al.(2006)]{2006ApJ...652..864H} Hopkins, P.~F., 
Somerville, R.~S., Hernquist, L., Cox, T.~J., Robertson, B., \& Li, Y.\ 
2006, \apj, 652, 864

\bibitem[{{Kaspi} {et~al.}(2005){Kaspi}, {Maoz}, {Netzer}, {Peterson}, {Vestergaard}, \& {Jannuzi}}]{Kas++05} {Kaspi}, S., {Maoz}, D., {Netzer}, H., {Peterson}, B.~M., {Vestergaard}, M., \& {Jannuzi}, B.~T. 2005, \apj, 629, 61

\bibitem[Kauffmann \& Haehnelt(2000)]{2000MNRAS.311..576K} Kauffmann, G., \& Haehnelt, M.\ 2000, \mnras, 311, 576

\bibitem[Lauer et al.(2007a)]{2007ApJ...662..808L} Lauer, T.~R., et al.\ 
2007a, \apj, 662, 808

\bibitem[Lauer et al.(2007b)]{2007arXiv0705.4103L} Lauer, T.~R., Tremaine, S., Richstone, D., \& Faber, S.~M.\ 2007b, \apj, 670, 249

\bibitem[McGill et al.(2008)]{2007arXiv0710.1839M} McGill, K.~L., Woo, J.-H., Treu, T., \& Malkan, M.~A.\ 2008, ApJ, 673, 703 (M08)

\bibitem[Onken et al.(2004)]{2004ApJ...615..645O} Onken, C.~A., et al. 2004, \apj, 615, 645

\bibitem[Peng et al.(2006)]{2006ApJ...649..616P} Peng, C.~Y., Impey, C.~D., Rix, H.-W., Kochanek, C.~S., Keeton, C.~R., Falco, E.~E., Leh{\'a}r, J., \& McLeod, B.~A.\ 2006, \apj, 649, 616

\bibitem[Robertson et al.(2006)]{2006ApJ...641...90R} Robertson, B., et al.\ 2006, \apj, 641, 90 

\bibitem[Salviander et al.(2007)]{2007ApJ...662..131S} Salviander, S., Shields, G.~A., Gebhardt, K., \& Bonning, E.~W.\ 2007, \apj, 662, 131 

\bibitem[Shen et al.(2008)]{2008AJ....135..928S} Shen, J., Vanden Berk, D.~E., Schneider, D.~P., \& Hall, P.~B.\ 2008, \aj, 135, 928

\bibitem[{{Shields} {et~al.}(2003){Shields}, {Gebhardt}, {Salviander}, {Wills},
  {Xie}, {Brotherton}, {Yuan}, \& {Dietrich}}]{Shi++03}
Shields, G.~A., et al. 2003, \apj, 583, 124

\bibitem[{{Tremaine} {et~al.}(2002){Tremaine}, {Gebhardt}, {Bender}, {Bower},
  {Dressler}, {Faber}, {Filippenko}, {Green}, {Grillmair}, {Ho}, {Kormendy},
  {Lauer}, {Magorrian}, {Pinkney}, \& {Richstone}}]{Trem++02}
{Tremaine}, S., et al. 2002, \apj, 574, 740

\bibitem[Treu et al. 2004]{treu04} Treu, T., Malkan, M., \& Blandford, R. D. 2004, ApJ, 615, L97 

\bibitem[Treu et al.(2005)]{2005ApJ...622L...5T} Treu, T., Ellis, R.~S., 
Liao, T.~X., \& van Dokkum, P.~G.\ 2005, \apjl, 622, L5

\bibitem[Treu et al.(2007)]{2007ApJ...667..117T} Treu, T., Woo, J.-H., Malkan, M.~A., \& Blandford, R.~D.\ 2007, \apj, 667, 117 (paper II)

\bibitem[{{Vestergaard} \& {Peterson}(2006)}]{V+P06}
{Vestergaard}, M. \& {Peterson}, B.~M. 2006, \apj, 641, 689



\bibitem[{{Walter} {et~al.}(2004){Walter}, {Carilli}, {Bertoldi}, {Menten},
  {Cox}, {Lo}, {Fan}, \& {Strauss}}]{Wal++04a}
{Walter}, F., et al. 2004, \apjl, 615, L17

\bibitem[Woo et al.(2004)]{2004ApJ...617..903W} Woo, J.-H., Urry, C.~M., Lira, P., van der Marel, R.~P., \& Maza, J.\ 2004, \apj, 617, 903
\bibitem[Woo et al.(2005)]{2005ApJ...631..762W} Woo, J.-H., Urry, C.~M., van der Marel, R.~P., Lira, P., \& Maza, J.\ 2005, \apj, 631, 762
\bibitem[Woo et al.(2006)]{2006ApJ...645..900W} Woo, J.-H., Treu, T., Malkan, M.~A., \& Blandford, R.~D.\ 2006, \apj, 645, 900 (paper I)

\bibitem[Woo(2008)]{2008arXiv0802.3705W} Woo, J.-H.\ 2008, AJ, in press (astroph/0802.3705)

\end{thebibliography}
\end{document}